\documentclass{kluwer}    

\usepackage[]{graphicx}

\begin{document}                                                                                   
\begin{article}
\begin{opening}         
\title{From Spirals to Low Surface Brightness galaxies}
\author{S. \surname{Boissier \email{boissier@ast.cam.ac.uk}}}
\institute{Institute of Astronomy, Cambridge, UK}
\author{D. \surname{Monnier Ragaigne}}
\institute{Observatoire de Paris, GEPI, Meudon, France}
\author{N.  \surname{Prantzos}}
\institute{Institut d'Astrophysique de Paris, Paris, France}
\author{W. \surname{van Driel}}
\author{C. \surname{Balkowski}}
\institute{Observatoire de Paris, GEPI, Meudon, France}

\runningauthor{Boissier et al.}
\runningtitle{From spirals to LSB galaxies}

\begin{abstract}
We show that simple models of the chemical and spectrophotometric evolution
of galaxies can be used to explore the properties of present-day
galaxies and especially the causes of the observed variety among disc galaxies.
We focus on the link between ``classical'' spirals and Low Surface Brightness galaxies.
\end{abstract}

\end{opening}

\section{Introduction}

In the last decades, a considerable amount of observations
has been collected on Low Surface Brightness galaxies 
(LSBs), which are characterised by a central surface brightness 
well below the Freeman disc centre value ($\mu_{B,0}$=21.65 mag arcsec$^{-2})$.
In a related contribution in the present Volume (Monnier Ragaigne et al.), 
the samples of LSBs used for comparison with our models are presented. 
LSBs are relevant for the study of the formation and evolution of 
galaxies in general, as well as for observational
cosmology, since some studies suggest that they may be 
responsible for a significant fraction of the high redshift
quasar absorbers in which gas densities and abundances can be measured
on cosmological scales (e.g. Boissier et al., 2002 and references therein).

In section \ref{seccomp}, we will compare the properties of the
observed samples with the predictions of models of their chemical and
spectro-photometric evolution based on reliable
models of spirals (presented in section \ref{secmod}).

\section{Models}

\label{secmod}

In the contribution by N. Prantzos, it is shown that the properties
of nearby spiral galaxies are in good agreement with a grid of simple models
of their chemical and spectrophotometric evolution. These models are based
on: 
i) a calibration on the Milky Way,  
ii) scaling relationships obtained in the context of the cold dark
matter theory (Mo et al., 1998), and
iii) an empirical calibration of infall time-scales: massive galaxies
accreting gas more rapidly than low mass galaxies (Boissier and Prantzos, 2000).

The second point is the most relevant for the study of LSBs, as the
scaling relations link the structural properties of the disc (scalelength and
central surface density) to those of the dark halo (circular velocity $V_C$ and spin 
parameter
$\lambda$): \\
$$
\label{equascal1}
R_d = R_{d,MW} \times \frac{V_C}{V_{C,MW}} \times \frac{\lambda}{\lambda_{MW}} \;  ; \; 
\Sigma_0=\Sigma_{0,MW} \times \frac{V_C}{V_{C,MW}} \times \left(\frac{\lambda}{\lambda_{MW}}\right)^{-2}.
$$ \\
The index $MW$  refers to the corresponding value in the Milky Way. 
The spin parameter $\lambda$ is a dimensionless quantity measuring the specific 
angular momentum of the dark halo.

\begin{figure}
\begin{center}
\begin{tabular} {c c }
&
\includegraphics[width=4.5cm,angle=-90]{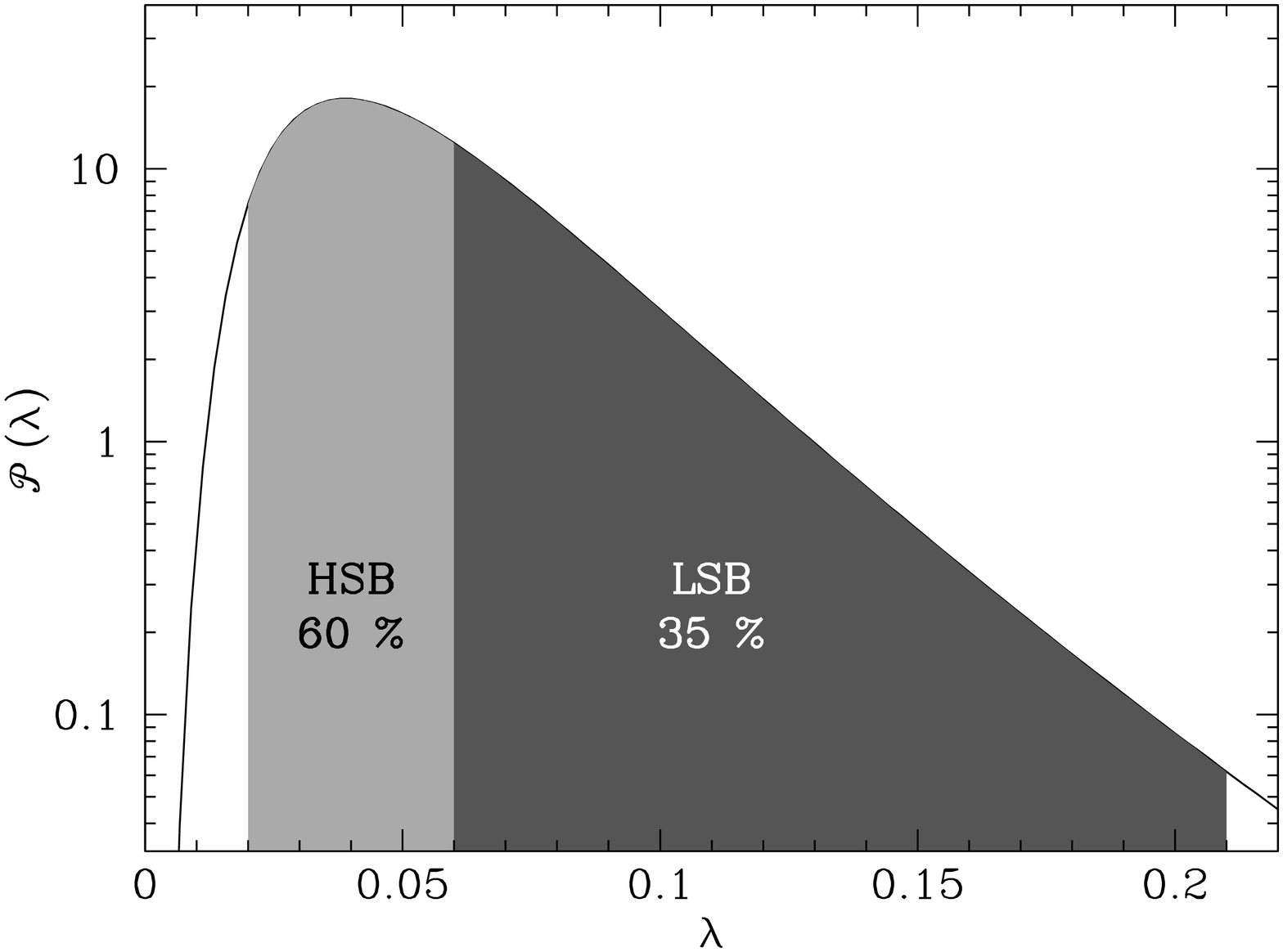} \vspace{-4.4cm}  \\
\includegraphics[width=5.cm]{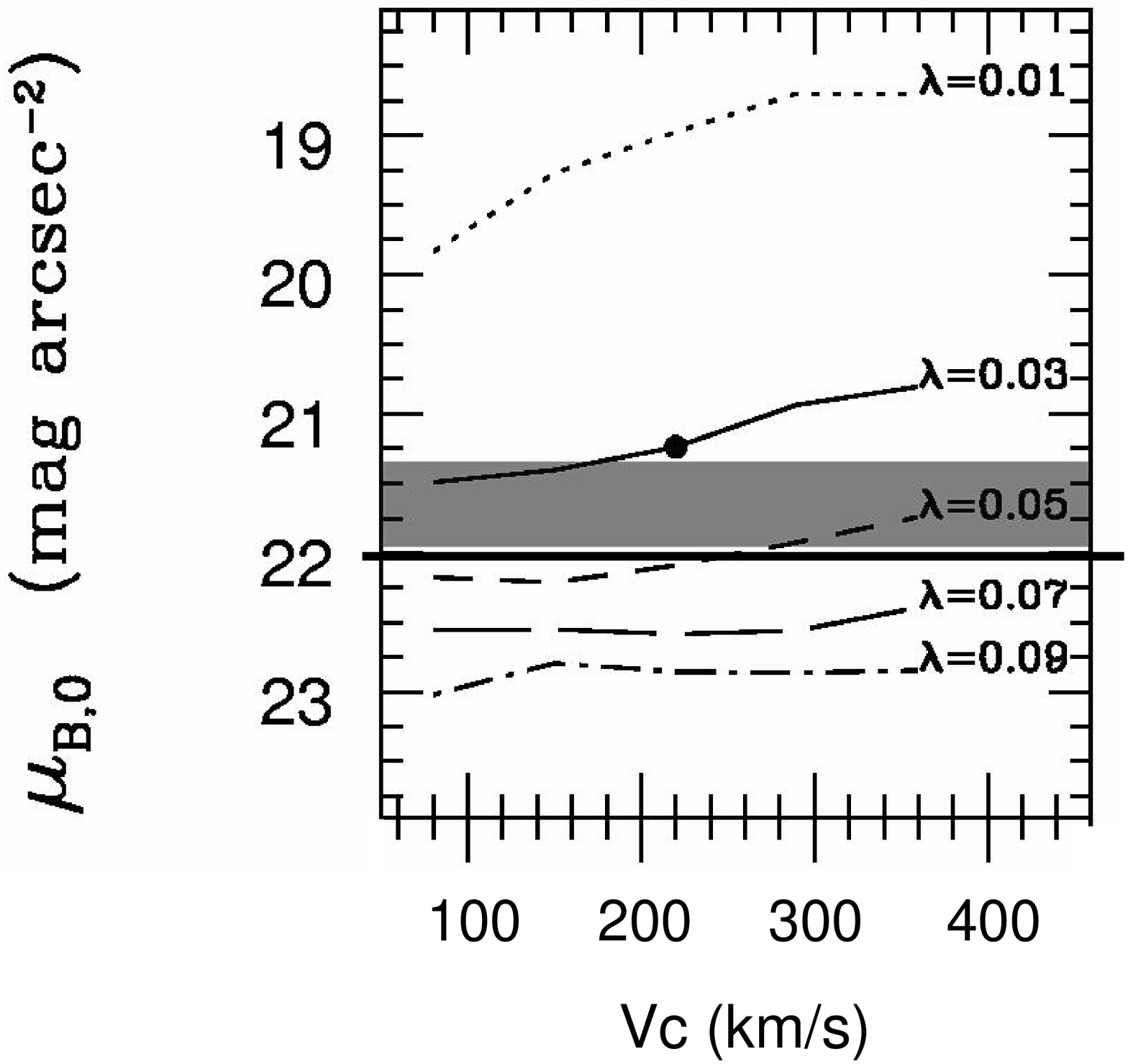} &
\end{tabular}
\end{center}
\vspace{-0.2cm}
\caption{\label{figspin}\underline{Left}: $B$-band central disc surface
brightness as a function of rotational velocity. Each line corresponds
to a different value of the spin parameter $\lambda$. The Freeman value, typical
of ``normal''spirals is indicated by the shaded area, while the limit of LSB
galaxies we adopted is indicated by the horizontal line. \underline{Right}:
distribution of the spin parameter $\lambda$. Models with $\lambda
\sim 0.04$ correspond to spirals with Freeman's surface brightness, 
models with $\lambda> 0.06$ to LSBs.}
\end{figure}

Dalcanton et al. (1997) already suggested that LSBs could be 
the large angular momentum equivalents of ``normal'' spirals,
an idea also used in the chemical evolution models of Jimenez et al. (1998).
Here, we will take advantage of the existence of a grid of models 
that are well calibrated for spiral galaxies.
Figure 1 shows the surface disc brightness obtained with
the spiral galaxy models, where, indeed, models with $\lambda>0.06$ are found to
be LSBs ($\mu_B>$ 22 mag arcsec$^{-2}$ ). 
Integrating over a distribution of the spin parameter, this
correspond to 35 \% of all galaxies (in number).
In the next section, we extend the models to larger values of the
spin parameter (0.07, 0.09, 0.15, 0.21) and compare the results with some of the
available observations for LSBs.

\section{A comparison of models with observations}

\label{seccomp}

\subsection{Central disc surface brightness and scalelength}

\begin{figure}
\includegraphics[angle=-90,width=12cm]{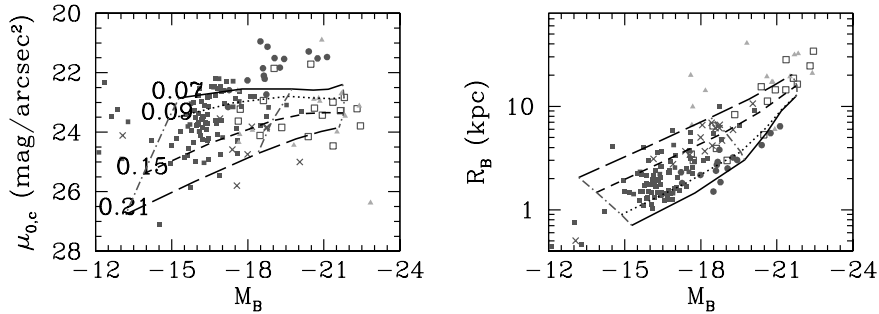}
\vspace{-4.7cm}
\caption{\label{figmurd}$B$-band central disc surface brightness (left) and
scalelength (right) as a function of the absolute magnitude of the disc
component for various observational samples (see Monnier Ragaigne et al., 
this Volume). The lines indicate models for LSB galaxies, i.e. with large 
spin parameter values, as indicated in the left panel.}
\end{figure}

The central disc surface brightness and the scalelength of the models
obtained with large values of the spin parameter are presented in
Figure 2 as a function of the absolute magnitude of the disc
component (all in the B band), for a number of samples of objects of
low and intermediate surface brightness, chosen to represent the wide 
variety among LSBs.  The comparison
between the two shows that these very simple models (and especially
the scaling relationships) provide a sound basis for the study of
LSBs.

LSBs could then well be disc galaxies with larger spin parameter than
spirals. As a consequence, for similar masses (and thus similar
absolute magnitudes, assuming that the mass-to-light ratio does not
vary drastically), the scalelengths of LSBs are expected to be larger
than for spirals. Indeed, the sample with intermediate surface
brightness is characterised by smaller scalelengths than the samples
with very low surface brightness galaxies.

\subsection{Chemistry}

Observational evidence on the chemical evolutionary state of LSBs is
relatively scarse. Their H\,{\sc i} mass-to-light ratios, on average larger than in spirals,
seems to indicate that LSBs are less evolved. This is
indeed what is found in the models, the star formation efficiency
being much smaller because of the smaller gas densities (see Monnier Ragaigne et al., 
this Volume).

For the same reasons, chemical abundances of heavy elements are
lower. In ``normal'' spirals, a clear relation between mass and metallicity is
observed (e.g. Zaritsky et al., 1994), which is reproduced in the models 
owing to the mass dependence of infall timescales. 
The observational situation is less clear for LSBs (see Figure 3).
However, contrary to the case of spirals, metallicity in LSBs is
measured in only a few H\,{\sc ii} regions located at various radii
in different galaxies: these data cannot be corrected for the unknown
underlying abundance gradients.  The abundances obtained in the LSB
models, weighted by a distribution of velocity, spin parameter and star formation rate 
(to take into account the need for massive stars to produce observable 
H\,{\sc ii} regions) are presented as the grey-scales in
Figure 3. A relatively large range of abundances is
expected, in rough agreement with the measurements.

\begin{tabular}{ p{5.4cm} p{5.4cm} }
\includegraphics[angle=-90,width=6cm]{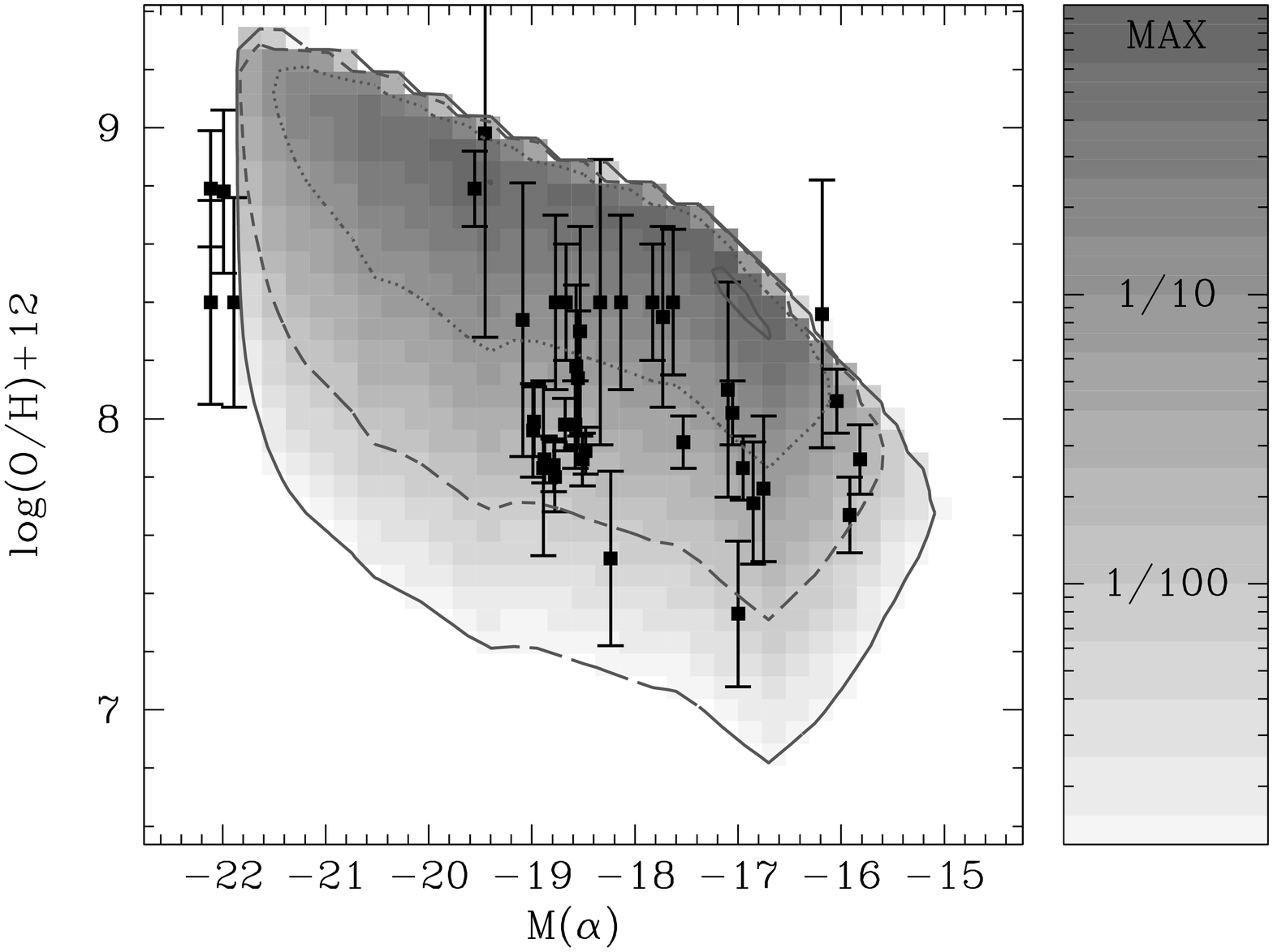} &
{\tabcapfont
\vspace{0.4cm} \textit{Figure 3.} Oxygen abundances in LSBs as a function of the
absolute blue magnitude of the disc component. Data are from Mc Gaugh
(1994). The grey-scale indicates expected values from the models,
weighted by a distribution of velocity, spin parameter, and star formation rate.} \\
\end{tabular}

\vspace*{-0.2cm}

\subsection{Colours}

The main difficulty for these ``simple'' models (with a smooth star
formation history) is their inability to explain the wide range of
colours observed (Monnier Ragaigne et al., this Volume). Adding to the
models starbursts and truncations of the star formation histories
seems a promising way to cure this problem (Boissier et al., in
preparation).

\end{article}

\end{document}